\documentclass[aps,pre,twocolumn,showpacs,showkeys,amsmath,amssymb,
groupedaddress]{revtex4}

\usepackage{graphicx}
\usepackage{times}
\newcommand{\prt}{\partial}
\newcommand{\prm}{\prime}
\newcommand{\mrm}{\mathrm}

\begin{document}

\preprint{}

\title{Analytical modelling of terminal properties in industrial growth}

\author{Arnabi Marjit}
\email{arnabi.marjit@rehau.com}
\affiliation{REHAU Polymers Pvt. Ltd. \\
Khed-Pabal Road, Holewadi, Rajgurunagar \\
Pune 410505, India}

\author{Sudipto Marjit}
\email{sudiptom@qaiindia.com}
\affiliation{QAI India Limited \\
1013-14A, Ansal Towers, 38, Nehru Place \\
New Delhi 110019, India}

\author{Arnab K. Ray}
\email{akr@iucaa.ernet.in}
\affiliation{Inter--University Centre for Astronomy and Astrophysics \\ 
Post Bag 4, Ganeshkhind, Pune University Campus \\ 
Pune 411007, India}

\date{\today}

\begin{abstract}
In this pedagogical study, carried out by adopting standard mathematical
methods of nonlinear dynamics, we have presented some simple 
analytical models to understand terminal behaviour in industrial growth. 
This issue has also been addressed from a dynamical systems perspective, 
with especial emphasis on the concept of the ``Balanced Scorecard". 
Our study 
enables us to make the general claim that although the fortunes of an
industrial organization can rise with exponential rapidity on relatively 
short time scales, its growth will ultimately and inevitably be saturated 
on long time scales by various factors which are nonlinear in character. 
We have mathematically demonstrated the likely occurrence of this feature 
under various possible 
circumstances, including the ``Red Ocean" and the ``Blue Ocean". Finally
and most importantly, our arguments and their associated  
mathematical modelling have received remarkable support from the growth 
pattern indicated by empirical data gathered from a well-recognized 
global company like {\it IBM}. 
\end{abstract}

\pacs{89.65.Gh, 05.45.-a, 87.23.Ge}
\keywords{Econophysics, business and management; Nonlinear dynamical systems;
Dynamics of social systems}

\maketitle

\section{Introduction}
\label{sec1}

Dynamic evolution is a ubiquitous attribute of all natural
phenomena. This should occasion no surprise, since any physical 
system that has had a beginning in time, under a given set of initial
conditions, has to make its expected passage through time. This is a
universal feature, from the
very largest scales, as it is with the Universe, whose evolution is
described by a plethora of dynamic cosmological models~\cite{dod}, 
to the very 
smallest, as it is with the growth of bacterial colonies~\cite{barstan}.

This kind of growth through
time, however, also has a terminal character about itself. Once again this
fact is not very difficult to intuit. On the largest scales the 
world can only be seen to be of a finite extent, constrained by 
physical boundaries, and so it could not possibly admit any physical
system within itself that will have the advantage of unbridled growth 
forever. Even if that were to apparently look so in the early stages of 
the evolution of a system, eventually there would be a saturation towards
some terminal limit. This message is driven home in no uncertain terms
by palpable facts that mountains can only grow up to a certain 
maximum height, 
or stars can remain as a stable physical configuration up to a certain 
limiting mass~\cite{weiss}. 

Mountains and stars, and generally speaking the Universe itself, all 
evolve much more slowly than events taking place on terrestrial and
human time scales, but their seemingly static and timeless appearance 
is nevertheless overridden 
ultimately by degeneracy and decay. A realization of this end state 
of all things though, can be had much more readily in biological systems. 
A quiet reflection on the extinction of most 
animal species to have dwelt on the Earth so far, is a sobering reminder 
of the mortality of all living beings, all of whom have to contend with 
the rude fact that they have to operate and thrive within a finite space,
characterized by parameters of finite size. As a result their environment
becomes necessarily competitive, with the resources at their disposal
being limited. This makes any indefinite and unconstrained growth 
an impossibility.  

While the dynamic behaviour of any natural phenomenon --- be it 
cosmological, biological, chemical or anything else of a likewise 
objective nature --- can be described by physical laws, enunciated 
in precise mathematical terms, it is also not difficult to 
recognize qualitatively 
similar evolutionary features in systems whose origins lie 
in the social functionings of the human species. The collective 
history of mankind is replete with examples of the dramatic rise 
of civilizations, all of which faded out eventually following a 
long decline through a twilight period. And so does it happen too
to political ideologies, socio-cultural mores and economic principles
(with the last of which is connected the fortunes of whole industries).

In this context we would like to pose the following question : 
Can the dynamics of a social system be described in equally 
precise and objective terms as we can do for a natural system? 
Addressing this question is the principal objective of this work, 
and to do so we have chosen to study the growth pattern exhibited
by an industrial organization. This choice has been inspired by the 
fact that many aspects of industrial growth lend themselves to 
reasonably objective analyses. The health of a company is to be
seen from the revenue that it is capable of generating, as well
as the extent of human resource that it is capable of engaging
in achieving its objectives. One could make a quantitative measure
of all these variables, and this makes it relatively easy to have 
a clear understanding of the growth pattern of an industry and
posit a mathematical model for it.

Our approach to these issues is predominantly based on the use of 
standard mathematical tools of nonlinear dynamics and dynamical
systems~\cite{stro,js99}. To explain the growth behaviour of an 
industrial 
organization, we have developed our theme by making a pedagogical
exposition of the relevance of various mathematical models of 
increasing complexity. We have found that even when an industrial
organization displays conspicuous growth in the early stages,
there is saturation of this growth towards a terminal end after 
the elapse of a certain scale of time. This has been cogently 
demonstrated in the three following cases: 

\begin{itemize}

\item
Growth in the presence of various inhibitive factors. These 
factors could be as varied as misalignment within an organization
itself, to external factors like competition with a rival, or 
the diverse socio-economic and ethnic milieu of the environment
in which a company might have to operate. 

\item 
Growth in which monopolistic control of the theatre of operations 
has been wrested. 

\item
Growth under conserved and constrained conditions, without any 
dissipative factors being active. 

\end{itemize} 
The conclusion in all the three cases above is the same: One way
or the other, as the system size (reflective of the scale of 
operations) begins to grow through the passage of time, a 
self-regulatory mechanism takes effective control over growth
and gradually drives the system towards a saturated terminal state. 

We have subjected our theory to empirical test. We have collected 
data pertaining to a representative industrial organization, and 
our prefence has been for a company which is global in nature, 
i.e. its presence is seen and felt all over the world. This choice 
has been with a design, because we would like to understand the global 
growth behaviour of a company, whose operating space is by definition 
on the largest available scale, and, therefore, the overall pattern 
of its growth  
would be free of local inhomogeneities. Speaking in terms of 
analogy, the Universe is also seen to be homogeneous and isotropic
on its largest scales~\cite{dod}, and the Earth itself, appearing 
uneven and
of infinite extent on local scales, presents a radically different
surface topology when viewed from outer space. And so it 
should be that on large scales a clear understanding could also 
be derived about the constrained feature of the space within which
an organization functions. 

To this end
we have made a study of the cumulative revenue generating capacity
and the growth of the human resouce strength of the multi-national
company, {\it IBM}. The data obtained were published on the {\it IBM}
website\footnote{\tt{http://www-03.ibm.com/ibm/history}} itself. 
It has been satisfying for us to note that, analyzed
according to our objectives and specifications, the {\it IBM} data
actually give a striking match with some of our mathematical models. 
Both the capacity for revenue generation and the human resource 
content of {\it IBM}, over a period of more than ninety years of
the existence of the company (this long period actually makes it
quite expedient for our study insofar as it is concerned with 
the growth of an industrial organization from its inception to 
its terminal stage), show an initial phase of exponential
growth, to be followed later by saturated growth towards a 
terminal end. 

Making a final important point is well in order. We have carried 
out a mathematical
analysis of the growth of an industrial organization using methods
that are the forte of specialists in nonlinear dynamics. However,
the area of application of these analytical tools is, after all, 
the professional domain of management specialists. 
The connection between these two vastly diverse disciplines need 
not be so obvious. So even as we have developed our mathematical 
models, we have made the necessary and the relevant connection to
practical matters of interest for those who are concerned with 
business administration and the management of industrial organizations. 
Some of these issues have dwelt on concepts ranging from 
the ``Balanced Scorecard"~\cite{kapnor1,kapnor2} 
to the ``Red Ocean" and the ``Blue Ocean"~\cite{kimmau}. 
We can now safely claim that in all similarly empirical cases where 
our theoretical methods may be applied, one could come remarkably 
close to constructing realistic and generic growth models, and that 
these models would also possess a substantial degree of predictive 
power.  

\section{Saturated growth in an adverse environment}
\label{sec2}

To appreciate what the recognizable quantitative features of terminal
growth behaviour might look like, let us start with a simple power-law
model described mathematically by 
\begin{equation}
\label{power}
\phi = a t^{\beta} , 
\end{equation}
in which $\phi$ is the variable whose growth we are interested in 
(it could be the revenue generating strength or the human resource content
of a company), $t$ is time, $a$ is a proportionality constant and 
$\beta$ is the growth exponent. The first derivative of $\phi$ with
respect to $t$ is given as 
\begin{equation}
\label{devpow}
\frac{{\mrm d} \phi}{{\mrm d}t} = a \beta t^{\beta - 1} . 
\end{equation} 
This equation indicates the rate of growth of $\phi$ through $t$. If
$\beta =1$, then quite obviously the growth will be perpetually linear,
while if $\beta > 1$, then $\phi$ will display an unbridled and 
divergent growth through time. For our purposes of understanding 
terminal growth, however, it will be necessary to consider the case
of $\beta < 1$. This will mean that with increasing time, $\phi$ will
indeed increase, but this rate of growth itself will be a decreasing 
function of time. And so as $t$ increases, the growth of $\phi$ will
continue to slow down. 

While all this will be quite true, 
the model given by Eq.~(\ref{power}) is also overly
simplistic. It does convey a basic notion of terminal growth, but it
fails to show that with increasing $t$, there will be a convergence 
towards a finite value of $\phi$. To achieve this constrained condition, 
it will be necessary to adopt a slightly more complex model. Let us 
choose an equation by which the early stages of growth will be 
conveniently linear (i.e. $\beta =1$). The slowing down will take 
place on later time scales. Such a situation can be easily described
by the equation, 
\begin{equation}
\label{stokes}
\frac{{\mrm d} \phi}{{\mrm d}t} = a - b \phi , 
\end{equation}
with both $a$ and $b$ being positive constants. In physics, this 
first-order linear ordinary differential equation is easily 
recognized as bearing the form of the equations which describe 
Stokes' law of terminal velocity of a particle falling through a 
viscous medium~\cite{landau}, and the viscoelastic deformation of 
rocks~\cite{lowrie}. Integration 
of Eq.~(\ref{stokes}) is trivial, and under the initial condition that
$\phi =0$ at $t=0$, we will get the analytical solution, 
\begin{equation}
\label{integstokes}
\phi = \frac{a}{b} \left(1 - e^{-bt}\right) . 
\end{equation}  

The long-time behaviour of this equation is worth a close look. When 
$t \longrightarrow \infty$, there is a convergence of $\phi$ towards
a finite value, $a/b$. We could also examine the features of this 
model on early time scales (quantified by the condition that 
$t \ll b^{-1}$). In this case, $e^{-bt} \simeq 1 - bt$, 
and, therefore, $\phi \simeq at$, very much linear, and very much 
as we would have wished it to be to demonstrate how rapid growth 
in the early stages can be slowed down towards a terminal end. 

\begin{figure}
\begin{center}
\includegraphics[scale=0.65, angle=0]{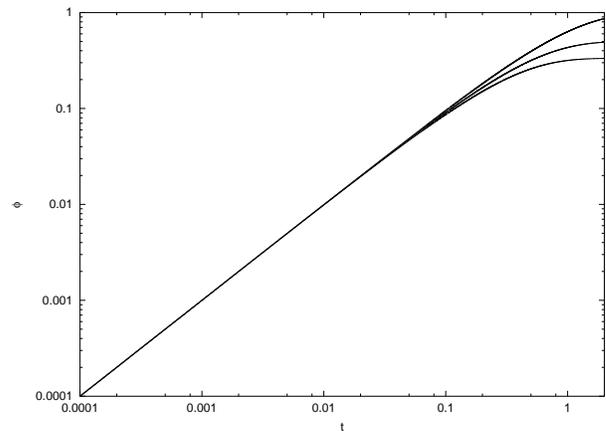}
\caption{\label{f1} \small{A logarithmic plot of $\phi$ against $t$,
to show how linear growth on small time scales can become saturated
towards a terminal value through an infinite passage of time, for a 
fixed value of $a$ but for three different values of $b$. All three 
solutions display the same early behaviour (for $a=1$), but the terminal 
stage is different for each case, depending on $b$. From the top to the 
bottom the three curves correspond to $b=1$, $2$ and $3$, respectively.}}
\end{center}
\end{figure}

In Fig.~\ref{f1} we have plotted Eq.~(\ref{integstokes}) for three
different values of $b$, at a fixed value of $a$. Both $\phi$ and 
$t$ have been plotted logarithmically for a better understanding of
the power-law behaviour in the early stages (should there be any),
and it is easy to see that with a progressive increase in the value
of $b$, the growth is saturated towards a terminal state at lesser
values of $\phi$, and sooner in $t$. 

At this stage it should be quite instructive to study the parameter
$b$ carefully. We can argue that the time scale for the onset of the
terminal character is given by $t \sim b^{-1}$. Intuitively this is 
just what it should be. The parameter $b$ quantifies everything that 
can retard growth, and so the time scale of terminal behaviour should,
qualitatively speaking, be in inverse proportion to it. That this is
indeed so is clearly portrayed in Fig.~\ref{f1}. Going back to the 
retarding term in Eq.~(\ref{stokes}), it can also be argued that 
the coupling of $b$ to $\phi$ will imply that near the terminal
state the growth of $\phi$ will be limited by its own inflated value. 

In the context of the growth of an industrial organization, we can 
identify various factors that may contribute collectively to $b$. 
These factors can be both economic and non-economic in nature. Some
may operate internally, while others can make their impact externally. 
A primary factor is the space within which an organization can be 
allowed to grow. If this space is constrained to be of a finite size
(as, practically speaking, it has to be), then, of course, terminal
behaviour becomes a distinct possibility. With its growth, an 
organization will gradually have to contend with the boundaries 
of the space within which it has to operate. This brings growth 
to a slow halt. This can be further aggravated by the presence 
of rival organizations competing for the same space. The last 
factor can become particularly acute when a miscalculation is made
in assessing future directions of growth vis-a-vis those of rival
organizations --- both the existing ones and the ones that 
might emerge in the future. 

By way of addressing all this,  
it might be argued that adopting an active policy based
on innovation, diversification or transformation could lead to more
sustained growth. However, there are some difficult practical issues
to be addressed in this situation too. When an organization transforms
itself or diversifies its core competency and objectives, it also 
runs the risk of rendering a sizeable fraction of its human resources
redundant. While this will undoubtedly leave an immediate and tangibly
adverse economic impact, the social impact of such measures could be
quite considerable too in variously incalculable ways. 

Changing political and social values can also be contributory factors
in retarding growth. To cite some specific examples in this regard, 
it is not difficult to perceive that a vigorous governmental pursuit 
of a policy of
disarmament can adversely affect the fortunes of the armaments 
industry, while a greater public awareness of health has nothing 
positive to offer to the tobacco industry. 

Increasing the base of operations of an organization by crossing national
boundaries may be seen as a lasting solution to problems related to
retarded growth. This will open greater opportunities for any organization
to operate and grow. While this may look like a self-evident truth,
once again, however, such measures will not be free of their associated
difficulties. When an organization
grows and steps beyond the boundaries of its country of origin, it
necessarily tends to stretch its existing resources. Quite
naturally this leads to a reduction in the intensity of its control 
mechanisms, and as a result there arises an early phase of misalignment 
that dampens the initial spurt in growth. Much conscious effort 
then needs to be undertaken to ensure functional coherence with the 
mother company.

Varied socio-cultural barriers across national frontiers are also 
never too easy to overcome, and adjusting the outlook and objectives
of an organization with the native ethnic milieu will, of necessity,
involve localizing the original character of the organization. This
will ultimately be reflected as localized misalignments in its global
functionings, with an attendant dampening of its desired extent of
growth. In this situation a significant degree of modification of the
policies and procedures of the company will have to be called for, 
keeping in mind the imperatives of alignment. Complete 
alignment may be prove to be elusive all the same, and hence,
there will be a certain extent of reconciliation to the inevitability
of misalignment in operations.

\begin{figure}
\begin{center}
\includegraphics[scale=0.65, angle=0]{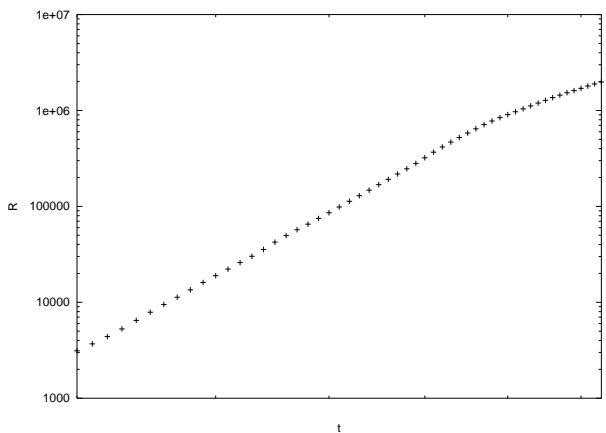}
\caption{\label{f2} \small{A logarithmic plot of the cumulative revenue
generation capacity, $R$, against time, $t$. The data show steady growth
from the middle of the $1950{\mrm s}$ till the present date. The later
stage of growth shows a distinct swing towards a terminal state.}}
\end{center}
\end{figure}

Misalignment could also result from acquisitions and mergers. 
Achieving an optimal state of consonance in core competency, style of 
functioning, and objectives between two hitherto different organizations, 
is almost always fraught with great difficulties. This does nothing 
to act favourably on the common growth of the newly-merged 
organizations. 

While we have been addressing these issues in general theoretical terms
so far, it should now also be appropriate to have an empirical insight
into some specific questions. We have already indicated that our theory
has been bolstered by a survey of the growth pattern of the 
multi-national company, {\it IBM}. This organization has been in 
existence in its presently known form for nearly a century. Besides
this, it has spread all over the globe. So on both of these counts, 
a company like {\it IBM} is ideally suited for our study. Data 
about its annual revenue generation and human resource strength, dating
from the year $1914$, have been published on the company website. Both
these indices of the health of the company have been useful for us. 
In Fig.~\ref{f2} we have presented a logarithmic plot of how the 
revenue generating ability of {\it IBM} has grown {\em cumulatively} 
since the year $1954$. The prescription of the cumulative growth needs
to be stressed upon here. For any evolving system its rate of growth 
is almost always a direct mathematical function of its current state. 
Frequently this dependence leads to an exponential growth pattern 
(something that, as we shall presently demonstrate, is quite relevant 
in this case also). Mindful of this fact, we have transformed the 
annual revenue generated by {\it IBM} into a cumulative revenue index,
$R$, which is measured in millions of dollars. This index has been 
plotted along the vertical axis of Fig.~\ref{f2},
with the horizontal axis measuring the logarithm of the time (scaled 
in years). The similarity of this plot with all the model solutions 
shown in Fig.~\ref{f1} is very apparent. The most interesting feature
is the diversion into a mode of reduced growth rate in the very
last decade of {\it IBM}. This is exactly the terminal growth phase
that we have depicted theoretically in Fig.~\ref{f1}. 

\begin{figure}
\begin{center}
\includegraphics[scale=0.65, angle=0]{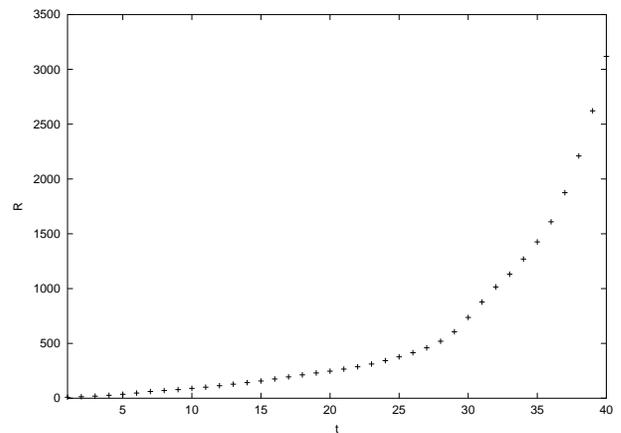}
\caption{\label{f3} \small{The early growth of the cumulative capacity for
revenue generation (spanning the first forty years of available data from
{\it IBM}) shows an exponential pattern.}}
\end{center}
\end{figure}

All this similarity between the empirical data and our theoretical
model is undoubtedly pleasing to note, but one could actually go much
beyond this. The model given by Eq.~(\ref{integstokes}) shows that the
early power-law type of behaviour will start developing right from the
inception of an organization. However, this is not exactly what the
{\it IBM} data indicate. For forty years since $1914$, the company has
registered a growth pattern in its cumulative revenue that is visibly
exponential, something that is quite evident from Fig.~\ref{f3}.

\begin{figure}
\begin{center}
\includegraphics[scale=0.65, angle=0]{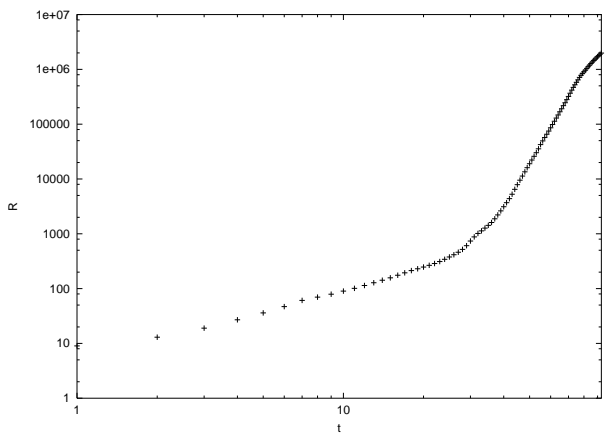}
\caption{\label{f4} \small{A logarithmic plot of the cumulative revenue
growth against time for nearly a century, shows that the early 
exponential growth of {\it IBM} has been saturated in the later stages.}}   
\end{center}
\end{figure}

\begin{figure}
\begin{center}
\includegraphics[scale=0.65, angle=0]{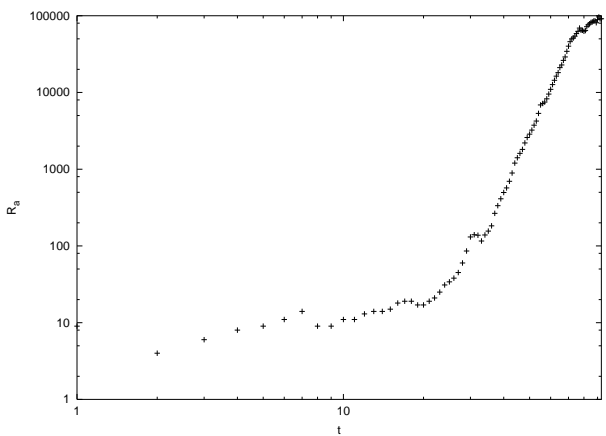}
\caption{\label{f5} \small{A logarithmic plot of the annual revenue,  
$R_{\mrm a}$, against time. It shows some fluctuations from
year to year, but the qualitative similarity with the trend indicated
by Fig.~\ref{f4} is evident. In fact the saturation towards a 
terminal state looks more prominent in this plot.}}
\end{center}
\end{figure}

Taking the information contained in Figs.~\ref{f2} \&~\ref{f3} together,
a clear impression of the overall fortunes of the company can be derived
from Fig.~\ref{f4}, which gives a logarithmic plot of the cumulative
revenue growth of {\it IBM} over nearly a century of its existence.
For comparison this may also be viewed against the plot in Fig.~\ref{f5},
which shows the growth in the {\em annual} revenue generated (in 
millions of dollars) every year. The trend in the two plots are largely 
identical, except for some understandable fluctuations in the latter.

There is a very interesting aspect to both Figs.~\ref{f4} \&~\ref{f5}.
Around the fortieth year of {\it IBM}, its growth seemed to have deviated 
noticeably from its early exponential rise. This could only have 
resulted from a stagnation that {\it IBM} might have faced in its native 
base. To overcome this impasse, it became necessary for {\it IBM} to
extend its core competency levels, as well as undergo a concerted 
process of geographical expansion. And indeed in the decade following 
the Second World War, {\it IBM} applied itself assiduously to the 
innovation and development of increasingly newer technology (the details 
of which are available on the {\it IBM} website), and to expanding its 
operational base overseas (for instance, {\it IBM} Israel was founded 
in $1950$ and {\it IBM} United Kingdom in $1951$). 

In the long run, however, even these measures proved insufficient in
sustaining growth, and at present {\it IBM} has globally lapsed into 
the terminal phase. This observation is in perfect consonance
with our earlier reasoning about how growth rates can become 
progressively diminished even after an organization employs a 
policy of expansion. First, it will be cramped for space once its 
very own size becomes comparable with the largest scales of available 
physical space, and secondly, various dissipative factors --- all 
of them presumably nonlinear in nature --- assume a very effective role
to throttle any further growth. 

\begin{figure}
\begin{center}
\includegraphics[scale=0.65, angle=0]{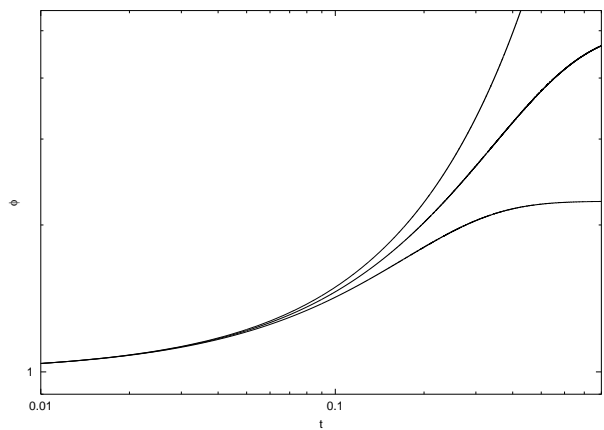}
\caption{\label{f6} \small{A logarithmic plot of $\phi$ against $t$,
to show how early exponential growth can be saturated on later time
scales by nonlinear factors. Under the initial condition, $\phi=1$
at $t=0$, the three curves from the top to the bottom correspond to
$\alpha = 0$, $1$ and $2$, respectively. Terminal features become more
pronounced for higher values of $\alpha$. All three curves have been
drawn for $a=5$ and $b=1$. }}
\end{center}
\end{figure}

Having taken cognizance of the inescapable empirical fact 
that after enjoying a period of early exponential rise (which implies
a beginning with a finite non-zero value at $t=0$, and then 
gradually capitalizing on this start), an industrial 
organization will have to ultimately suffer its growth being stifled,
we are in consequence led to the conclusion that our relatively simple 
model, based on the assumption of an early linear growth, as described 
by Eq.~(\ref{stokes}), will not entirely suffice 
for our purposes of building an analytical model for terminal 
behaviour. So we will now have to posit a nonlinear mathematical 
model that will represent a significant improvement over the simple 
linear differential equation described by Eq.~(\ref{stokes}). 
According to this necessity, as a simple extension of the linear
model, we put forward a nonlinear growth model, suited for all 
time scales as 
\begin{equation}
\label{nonlin}
\frac{{\mrm d}\phi}{{\mrm d}t} =\phi \left(a-b\phi^\alpha\right) , 
\end{equation}
in which $\alpha$ is an exponent, which, for the sake of simplicity, 
is restricted to be either zero or a positive integer. We also impose
the requirement that $a > b$, with both of these parameters being 
positive. 

Since Eq.~(\ref{nonlin}) is a first-order differential equation in 
time, we will need only one initial condition to fix the first 
integral of the equation. To this end we require that 
$\phi = \phi_0$ at $t=0$, for any value of $\alpha$. Let us first
consider the case of $\alpha =0$. This will immediately reduce 
Eq.~(\ref{nonlin}) to a linear differential equation, whose integral
solution will read as 
\begin{equation}
\label{alpha0}
\phi = \phi_0 e^{(a-b)t} . 
\end{equation} 
This solution, of course, represents an unbridled exponential growth
that can be sustained perpetually. The topmost curve in Fig.~\ref{f6}
depicts this behaviour. 

More interesting properties will result from Eq.~(\ref{nonlin}) when
$\alpha > 0$. This case will imply that the exponential growth exhibited
by Eq.~(\ref{alpha0}) will be saturated by a nonlinear damping term.
For $\alpha =1$, the damping term will be of the second order of
nonlinearity. The corresponding integral solution, obtained by the
method of partial fractions, will be
\begin{equation}
\label{alpha1}
\phi = \frac{ac}{bc + e^{-at}} ,
\end{equation}
in which $c = \phi_0 (a - b \phi_0)^{-1}$. This solution is represented
by the second curve from the top in Fig.~\ref{f6}. The  
profile of this theoretical solution has much closeness with the two 
empirical growth profiles in Figs.~\ref{f4} \&~\ref{f5}, respectively. 

Prescribing $\alpha = 2$ will raise the damping term in Eq.~(\ref{nonlin})
to the third order of nonlinearity, and the integral solution will 
accordingly be obtained as 
\begin{equation}
\label{alpha2}
\phi = \frac{ck}{\sqrt{c^2 + e^{-2at}}} , 
\end{equation} 
with $c = \phi_0(k^2 - \phi_0^2)^{-1/2}$ and $k^2 = a/b$. This solution
has been plotted as the lowermost curve in Fig.~\ref{f6}. What we 
can immediately conclude from the behaviour of the three curves in 
Fig.~\ref{f6} is that with progressively higher orders of nonlinearity,  
the damping of the growth of $\phi$ is saturated towards the terminal 
phase with greater promptness. And much more importantly for our study,
we can also appreciate that the characteristic of nonlinear saturation 
of exponential growth is very much in keeping with what the {\it IBM}
data indicate regarding growth in the revenue generating abilities 
of an industrial organization. When $\alpha >0$, the terminal value 
of $\phi$ will be given by $(a/b)^{1/\alpha}$, and the onset 
of nonlinear damping will occur when the magnitude of $\phi$ becomes 
comparable with this terminal value. It is not difficult to see that 
near this critical value, any further growth of $\phi$ would be inhibited
by its own size. For the specific cases of Eqs.~(\ref{alpha1}) 
\&~(\ref{alpha2}) this contention is easily verified for 
$t \longrightarrow \infty$. In the opposite limit of 
$t \longrightarrow 0$, it is equally apparent that both the damped
solutions will merge into an exponential pattern. 

Finally, we note that the value of $\alpha$ has a bearing on 
the symmetry of Eq.~(\ref{nonlin}). We have constrained $\alpha$ to
belong to the set of positive integers. When $\alpha$ is an even 
number, Eq.~(\ref{nonlin}) retains an invariant form under the 
transformation, $\phi \longrightarrow - \phi$. On the other hand, 
when $\alpha$ is an odd number, this invariance breaks down. In physics,
this argument, based on considerations of symmetry, is crucial for 
the precise choice of equations governing various nonlinear growth 
phenomena~\cite{barstan}. 
For the specific instances of $\alpha =1$ and $\alpha =2$
in Eq.~(\ref{nonlin}), there are many analogous equations
in nonlinear dynamics. To cite one such example with regard to the 
former case, there is the law of mass action of chemical 
kinetics~\cite{stro}.
With regard to the latter, there is the nonlinear saturation of the
amplitude of a standing wave in super-critical flows, in the fluid
dynamical problem of the hydraulic jump~\cite{rbh07}. 

Our industrial case study may not need to invoke such intricate 
details in making a correct choice of a value for $\alpha$.  
All the relevant industrial variables in our study are always 
measurably positive 
quantities, and so in a qualitative sense any integral exponent 
(odd or even) should be effective to bring about terminal growth.
The simplest case, of course, is when $\alpha =1$. 
So in concluding this discussion we can now claim that industrial
growth patterns can be described on early time scales by an 
exponential law, which is later driven towards a limiting value 
by various retarding factors, all of which are nonlinear in character. 

\section{The Balanced Scorecard : A dynamical systems perspective}
\label{sec3}

One crucial challenge that organizations face as
they mature and go beyond a critical size, is how to balance their
key stakeholder expectations. In this context it will be topical
to mention the paradigm of alignment (on which we have already dwelt 
at some length in Section~\ref{sec2}), using the ``Balanced Scorecard"
methodology. This was first proposed by 
Kaplan \& Norton~\cite{kapnor2}, 
who maintain that organizations
need to align the value propositions at the enterprise level, the 
business unit level and the support function level to create sustained 
organizational growth, in addition to balancing the financial and 
non-financial goals of the organization.

In point of fact, Kaplan \& Norton define a radically new perspective 
to the concept of alignment, and they strongly advocate that alignment 
should be instituted as a continuous process, rather than as a one-time 
annual ritual of goal setting and performance appraisal.  
An organization is a dynamic entity in a dynamically evolving
macroeconomic environment where change is continuously to be expected. 
Any strategy and its implementation must, therefore, evolve accordingly. 
Otherwise an organization, happily aligned at one point of time, will 
soon become misaligned. This is somewhat reminiscent of the onset and
growth of disorder (measured by entropy) in a thermodynamical system 
that is left to its own means~\cite{reif}. By the same token and as a
corollary of this physical analogy, the  
proposition of Kaplan \& Norton implies that keeping an organization 
on its sustained growth path, will necessitate the management of 
its process of alignment through active intervention. 

Frequently, however, it happens that even a very regular monitoring 
of industrial 
growth by means of one particular index only, may actually convey no 
more than a partial notion
of the true state of affairs. In such a situation a more balanced
view could be obtained by accounting for the behaviour of some
other relevant variables. Indeed, the concept of the ``Balanced Scorecard"
is premised on this very principle~\cite{kapnor1}.
Therefore, charting the individual properties of various key variables,
and then analyzing them in conjunction with one another, can enable
us to build a more comprehensive picture. The variables involved 
can be diverse in nature, ranging from the finances of an organization
to its human and technological resources and to the market within 
which the organization might be operating. The growth rate of any 
one of these variables may have a correlated functional dependence 
on the current state of all the pertinent variables.  

So if we are to study an industrial organization whose current state 
is defined completely by a set of $n$ variables, then the growth rate 
of the $i$-th. variable, $\phi_i$, can be described as, 
\begin{equation}
\label{gendyn}
\frac{{\mrm d}\phi_i}{{\mrm d} t} = \Phi_j(\{\phi_i\}) , 
\end{equation}
with $\Phi_j$ being a general function of all the variables in the set. 
The whole set can be expressed explicitly by making both $i$ and $j$
run from $1$ to $n$. In this situation we will have a set of $n$ 
first-order 
differential equations, with each one of them being coupled to all
the others. This entire set will form a first-order dynamical 
system, and for the sake of simplicity we require  
this dynamical system to be autonomous~\cite{stro,js99}.  

\begin{figure}
\begin{center}
\includegraphics[scale=0.65, angle=0]{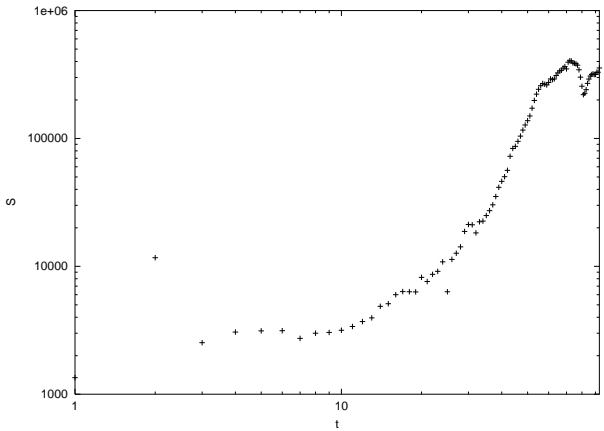}
\caption{\label{f7} \small{A logarithmic plot of the growth of 
human resource content, $S$, against $t$. The growth
indicated in this plot is compatible with the growth of $R$ and
$R_{\mrm a}$, as shown in Figs.~\ref{f4} \&~\ref{f5}, respectively. 
This implies that the growth of $R$ and $S$ are coupled to each other
and are correlated.}} 
\end{center}
\end{figure}

We have already carried out an extensive study of the growth of 
{\it IBM}, by following the revenue (both cumulative and annual) 
that the company has been generating over the years. Alongside 
this, we would now like to study the behaviour of another variable
--- the human resource content, $S$. Its growth against time has been
shown in Fig.~\ref{f7}. It is easy to discern that the general 
qualitative pattern of the growth of $S$ has been in tandem with 
the growth of $R$ or $R_{\mrm a}$, as Figs.~\ref{f4} \& \ref{f5}
show them. Therefore, it stands to reason that the growth of both 
$R$ (we prefer this variable to $R_{\mrm a}$) and $S$ are dynamically
coupled to each other in the form 
\begin{eqnarray}
\label{dynsys} 
\frac{{\mrm d}R}{{\mrm d}t} &=& {\rho}\left(R,S\right) \nonumber \\
\frac{{\mrm d}S}{{\mrm d}t} &=& {\sigma}\left(R,S\right) . 
\end{eqnarray} 
The two foregoing expressions should occasion no mystery. If an industrial
organization generates enough revenue, it should become financially
viable for it to maintain a sizeable human resource pool. On the other
hand, for a healthy organization, the human resource strength will 
translate into a greater ability to generate revenue. And in this 
manner both the revenue and the human resource content of an organization
will sustain the growth of each other. The coupled set involving $R$ 
and $S$ in Eqs.~(\ref{dynsys}) simply states this fact in mathematical 
terms. 

The equilibrium condition is obtained when both the derivatives on 
the left hand side of Eqs.~(\ref{dynsys}) vanish simultaneously, 
i.e. ${\mrm d}R/{\mrm d}t = {\mrm d}S/{\mrm d}t =0$. The corresponding
coordinates in the $S$---$R$ plane may be labelled 
$(S_{\mrm c},R_{\mrm c})$. Since the terminal state implies the 
cessation of all growth in time (i.e. all derivatives with respect 
to time will vanish), we are now in a position to argue that the 
equilibrium state in the $S$---$R$ plane actually represents a terminal
state in real time growth.  

Some general deductions can now be made about the nature of the 
equilibrium state, with the help of dynamical systems 
theory~\cite{stro,js99}. The two coupled equations, given by 
Eqs.~(\ref{dynsys}), will, in the most general sense, 
be nonlinear. A linearization
treatment on them could be carried out by applying small perturbations
on $R$ and $S$ about their equilibrium state values. The perturbation
scheme will be $R=R_{\mrm c}+R^{\prm}$ and $S=S_{\mrm c}+S^{\prm}$. 
This will allow us to express a coupled set of linearized equations as
\begin{eqnarray}
\label{linsys}
\frac{{\mrm d}R^{\prm}}{{\mrm d}t} &=& {\mathcal A}R^{\prm} 
+ {\mathcal B}S^{\prm} \nonumber \\
\frac{{\mrm d}S^{\prm}}{{\mrm d}t} &=& {\mathcal C}R^{\prm}
+ {\mathcal D}S^{\prm} , 
\end{eqnarray}
in which 
\begin{displaymath}
\label{coeffs}
{\mathcal A}= \frac{\prt {\rho}}{\prt R}
\bigg{\vert}_{R_{\mrm c}}, \, \, 
{\mathcal B}= \frac{\prt {\rho}}{\prt S}
\bigg{\vert}_{S_{\mrm c}}, \, \,
{\mathcal C}= \frac{\prt {\sigma}}{\prt R}
\bigg{\vert}_{R_{\mrm c}}, \, \,
{\mathcal D}= \frac{\prt {\sigma}}{\prt S}
\bigg{\vert}_{S_{\mrm c}} . 
\end{displaymath} 

Solutions of the form $R^{\prm} \sim e^{\lambda t}$ and 
$S^{\prm} \sim e^{\lambda t}$ will enable us to derive the 
eigenvalues, $\lambda$, of the stability matrix implied by
Eqs.~(\ref{linsys}), as 
\begin{equation}
\label{eigen}
\lambda = \frac{1}{2}
\left[\left({\mathcal A}+{\mathcal D}\right)\pm
\sqrt{\left({\mathcal A}+{\mathcal D}\right)^2 - 
4\left({\mathcal A}{\mathcal D} - 
{\mathcal B}{\mathcal C}\right)}\right] . 
\end{equation} 

The exact determination of the values of the two roots of $\lambda$ 
will impart a clear idea about the nature of the equilibrium state,
which can either be a saddle point or a node or a focus~\cite{stro,js99}.
The last case will necessarily mean an oscillatory nature in the 
growth of both $R$ and $S$~\cite{stro,js99}. If we go back to 
Figs.~\ref{f4} \&~\ref{f7}, we will notice that the respective growth 
patterns of {\em both} $R$ and $S$ have, on the other hand, been largely 
monotonic in nature (except for a noticeable fluctuation in $S$ at 
large values of $t$). So an immediate conclusion that follows is 
that the equilibrium state is very likely an unstable  
node~\cite{stro,js99}, and this will correspond mathematically 
to $\lambda$ having two real positive roots. Practically speaking, 
this is what we should expect entirely. For an industrial 
organization it is not conceivable that while there is growth in
one variable, there will be decay in the other. Both will have to
grow in close mutual association, and, if our dynamical systems argument 
is anything to go by, both will make the approach towards the terminal
state simultaneously. One variable cannot behave completely 
independently of the other. 

There is, however, a problem
inherent in all this. Since both the equilibrium states in $R$ and
$S$ (which are also the terminal states of these two variables) 
represent a maximum possible growth in time, the whole configuration 
is unstable. Slight fluctuations can bring about radical and, very
likely, unwelcome changes in this configuration.  
In this situation, a conventional approach to 
achieving high-growth performance (as, for instance, by arbitrarily
increasing or reducing the human resource strength) will not only
be merely cosmetic, but can actually be counter-productive. 
A truly effective remedy should be founded on a more imaginative and 
innovative approach. 

\section{Becalmed in the ``Blue Ocean" : Dynamic evolution 
towards the terminal state} 
\label{sec4}

It will now be both very relevant and very important to bring in here 
the concept of the ``Blue Ocean Strategy", introduced and popularized 
by Kim \& Mauborgne~\cite{kimmau}, who, for the very first time,
posited a contrary approach to strategy 
development. They colourfully refer to the traditional 
process of strategic planning as the ``Red Ocean", which, being 
based on competition, will inevitably lead to the shedding of 
``blood". In a ``Red Ocean" paradigm, a traditional organization focuses 
on competing against rivals and outmanoeuvring them. However, with 
supply exceeding demand in matured industries and markets, competing 
successfully for a share of contracting market spaces will not be 
sufficient to sustain high-growth performance. 

In a ``Blue Ocean" strategic paradigm of growth, on the other hand, 
organizations concentrate on ``value innovation" --- doing something beyond 
the conventional approach and not using competition as the benchmark. 
Instead the focus is on making competition irrelevant by creating a 
leap in value for customers and the organization, and thereby 
continuing to open up new and uncontested market spaces. This makes 
it apparently self-evident that if organizations successfully continue 
to create their niches of ``Blue Ocean", they may extend their period 
of exponential growth or defer their phase of terminal growth. 
We make an attempt to address these issues now through our particular
approach. 

First let us consider the ``Red Ocean" scenario. In doing so it 
will be instructive for us to invoke an analogy where the possibility 
of bloodshed, in its truest sense, is very much real --- that of 
two species of animals,
with population sizes, $\phi_1$ and $\phi_2$, competing for the
control of a common food resource. This condition can be expressed
in terms of a coupled autonomous nonlinear dynamical system as 
\begin{eqnarray}
\label{volterra}
\frac{{\mrm d}\phi_1}{{\mrm d}t} &=& \left[a_1 - d_1\left(b\phi_1 
+ c\phi_2 \right)\right]\phi_1 \nonumber \\
\frac{{\mrm d}\phi_2}{{\mrm d}t} &=& \left[a_2 - d_2\left(b\phi_1
+ c\phi_2 \right)\right]\phi_2 , 
\end{eqnarray} 
from which it is possible to show that if $a_1 d_2 > a_2 d_1$, then
the population size given by $\phi_2$ dies out, while the population 
size, $\phi_1$, approaches a limiting value~\cite{js99}. 
This is one manifestation
of Volterra's Exclusion Principle~\cite{js99}. Applied to a situation
where there is competition between two (or more) 
rival organizations for the 
control of a common market space, even the wresting of monopolistic
control by any one organization will not ensure its high-growth performance
in the long run. So there will be no protracted savouring of the fruits
of absolute victory over all rivals in the ``Red Ocean". This, of 
course, provides robust justification for adopting the ``Blue Ocean
Strategy". However, we shall now argue that even in doing so, one has 
to face the stark reality of growth being terminated ultimately. 

By its very definition the ``Blue Ocean" conjures up an image of 
vast openness. But does this actually imply an infinitely available
space for growth? In Section~\ref{sec2} we presented some models 
in which growth was being retarded by various dissipative factors,
either linear, or, more realistically, nonlinear. In a discussion
on what brings growth to a halt, we alluded to the finite size of
a system being one of the reasons that can constrain growth. This 
is something that we shall take up in greater detail here, with 
the help of some pedagogical models. 

Let us say that the ``Blue Ocean" defines the maximum available 
space in a niche that an organization has defined for its own growth.  
As it was before, the state of the organization is indicated by the 
variable,
$\phi$. However, unlike what we did before, requiring $\phi$ to be 
dependent on $t$ only, this variable will now also be a function 
of the spatial dimension, $x$, that it occupies inside the ``Blue 
Ocean" (which, for simplicity, we treat as one-dimensional). To 
enunciate this mathematically, we will 
write $\phi \equiv \phi(x,t)$. Let us also assume that none
of the usual retarding factors is operating against the growth 
of $\phi$, i.e. the system is a conservative one. 

Considering first a linear model of growth through the ``Blue Ocean", 
we take up the diffusion equation, 
\begin{equation} 
\label{diffu}
\frac{\prt \phi}{\prt t} = \Delta \frac{{\prt}^2 \phi}{\prt x^2} . 
\end{equation} 
Under the initial condition, $\phi(x,0) = \delta (x)$, the Dirac
delta function, we can arrive at the solution, 
\begin{equation}
\label{soldiffu}
\phi(x,t) = \frac{1}{\sqrt{4 \pi \Delta t}} \exp\left(- 
\frac{x^2}{4 \Delta t} \right) . 
\end{equation} 
This is the ``point source solution" of the diffusion equation~\cite{kbo}. 
It means that if $\phi$ were to start initially with an abrupt and high
spurt at a given point, it would subsequently diffuse (through the ``Blue
Ocean", as it were) according to the Gaussian distribution law 
given by Eq.~(\ref{soldiffu}). 
What emerges is that left to its own devices, there will be no growth 
whatsoever for $\phi$, but only its global redistribution, as 
the diffusion process progresses gradually. 
This is hardly surprising, because the diffusion model is a linear 
model, with nothing to drive the growth of $\phi$. 

However, this is not to say that no growth could ever be described by
a linear equation. For instance, the addition of a random noise term,
dependent on both $x$ and $t$, on the right hand side of Eq.~(\ref{diffu})
will alter this equation to the form of the Edwards-Wilkinson 
equation~\cite{barstan,ew}, which is a linear equation that is 
regularly invoked to follow the growth of an interface by the random
deposition of particles with surface relaxation~\cite{barstan}. As
a result of this relaxation, a correlated surface growth will be 
obtained through the redistribution of matter at the interface, after
starting from flat initial conditions. 

Inclusion of nonlinearity, on the other hand, generalizes the 
Edwards-Wilkinson equation to the Kardar-Parisi-Zhang 
equation~\cite{barstan,kpz}. The role of nonlinearity is crucial in
the latter case. It generates the growth of an interface by {\em adding}
material, as opposed to {\em reorganizing} the interface through 
redistribution of matter. So this is a situation where nonlinear 
effects play a more influential role in growth. Having said this, 
we must also point out that in keeping with the principle of our work,
growth models described both by the Edwards-Wilkinson and the 
Kardar-Parisi-Zhang equations come to a saturated end at a limiting
scale of length~\cite{barstan}. 

With regard to our specific objective of understanding how nonlinear
effects may improve on a linearized model, we now devise a simple 
first-order nonlinear mathematical model by which the growth of
the field, $\phi(x,t)$, will be described. We write it as 
\begin{equation}
\label{phieuler}
\frac{\prt \phi}{\prt t} + \phi \frac{\prt \phi}{\prt x} = 
- V^{\prm}(x) . 
\end{equation} 
The foregoing equation bears the form of pressure-free Euler's 
equation~\cite{landau}, which finds its use in the study of 
astrophysical gas dynamics~\cite{shu}. The term on the right hand
side gives the force that drives the growth of $\phi$. This force
is expressed as the negative gradient (the prime refers to a 
derivative with respect to $x$) of a potential function, $V(x)$. 

A solution of Eq.~(\ref{phieuler}) can be obtained by the method 
of characteristics~\cite{ld97}, which will entail our writing 
\begin{equation}
\label{charac}
\frac{{\mrm d}t}{1} = \frac{{\mrm d}x}{\phi} = 
\frac{{\mrm d}\phi}{-V^{\prm}(x)} . 
\end{equation}
The solution of the static ${\mrm d}\phi/{\mrm d}x$ equation is easy to 
find. It is 
\begin{equation}
\label{statchar}
\frac{\phi^2}{2} + V(x) = \frac{c^2}{2} , 
\end{equation} 
where $c$ is an integration constant, which, for this integration,
can in general be a function of $t$. It shall be presently demonstrated
that it actually vanishes terminally in time. The result of 
Eq.~(\ref{statchar}) can be used to find the solution of 
the ${\mrm d}x/{\mrm d}t$ equation. We can write this as 
\begin{equation}
\label{timechar} 
t = \pm \int \frac{{\mrm d} x}{\sqrt{c^2 - 2 V(x)}} , 
\end{equation}
from which, for our subsequent analysis, we will choose the positive 
sign of the square root by the 
self-evident physical criterion that $t > 0$. 

Prescribing a functional form of $V(x)$ will now be necessary. 
We have some general expectations about the driving force. It will
enable our system to stay well-knit, but at the same time will 
allow it to expand gently. This is certainly to be much
preferred to a violently fissiparous outward disintegration (which 
assuredly is never a desirable mode of industrial expansion). In
a very large ``Blue Ocean", we should also expect the 
influence of the driving force to vanish at the outer boundary,
from a simple requirement of well-behaved boundary conditions. 
These criteria can be mathematically fulfilled by various 
prescriptions for $V(x)$, but here we choose a very familiar
and most elementary expression, that of the Newtonian potential,
$V(x) = -1/x$. The negative sign in the potential keeps our
system self-contained, while the inverse law satisfies our required
boundary condition. In studies of astrophysical accretion, the 
``pressure-free" spherically symmetric infall of gas has been 
studied in great detail by using this potential~\cite{rb02,rr07}.  

\begin{figure}
\begin{center}
\includegraphics[scale=0.65, angle=0]{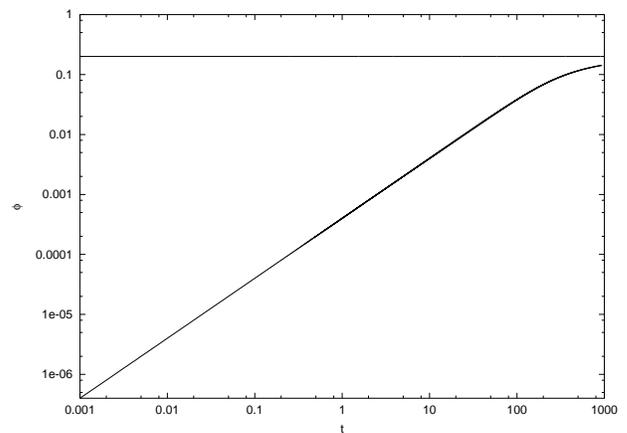}
\caption{\label{f8} \small{A logarithmic plot of the temporal
evolution of $\phi$ through $t$, at a fixed value of $x=50$.
The slope of this logarithmic plot shows that the early stages
of the evolution follow a linear power law. Deviation from this trend
sets in later. The horizontal line at the top shows the terminal
value that $\phi$ may attain as $t\longrightarrow \infty$.}}
\end{center}
\end{figure}

With the Newtonian potential defined for $V(x)$, it will now
become possible to derive an exact analytical solution for 
Eq.~(\ref{timechar}) as 
\begin{equation}
\label{timecharnewt}
c \phi x - \ln \left[x\left(\frac{\phi}{c} + 1\right)^2 \right]
- c^3 t = \tilde{c} . 
\end{equation} 
The general solution of Eq.~(\ref{phieuler}) will be delivered 
under the condition $f(\tilde{c}) = c^2/2$, with the functional
form of $f$ having to be determined by the initial condition 
under which an organization begins its voyage in the ``Blue 
Ocean". Realistically speaking, one suitable initial condition 
can be the globally flat condition, $\phi(x,0) =0$. 
At subsequent times the evolution of $\phi$ 
will be driven by our chosen force, and the resultant analytical
relation for the growth of $\phi$ will be given by 
\begin{equation}
\label{finsol}
\phi^2 - \frac{2}{x} \left[ 1 - \left(\frac{\phi}{c} + 1\right)^{-2} 
\exp \left(c \phi x - c^3 t \right) \right] = 0 . 
\end{equation} 
It can now be seen that for $t \longrightarrow \infty$, we will 
have the solution, 
\begin{equation}
\label{asymp1}
\phi^2 = \frac{2}{x} ,
\end{equation} 
i.e. through an infinite passage of time, $\phi$ can only grow
up to a limiting value. This limit will be determined by the 
spatial dimensions within which an organization will be operating. 
We make a graphical representation of our arguments in Fig.~\ref{f8}
by applying finite differencing techniques on Eq.~(\ref{phieuler}). 
It can be seen in this plot that at a given spatial position, the 
temporal evolution of the $\phi$ field will reach a terminal value
after an infinite elapse of time. This property will be independent
of the chosen spatial scale. 

If, suppose, we choose another initial condition, $\phi(x,0)=\phi_0$,
which is a constant non-zero initial value of $\phi$, then we can 
show that for $t \longrightarrow \infty$, we will have $\phi$ 
converging according to, 
\begin{equation}
\label{asymp2}
\phi^2 = \phi_0^2 + \frac{2}{x} ,
\end{equation} 
which underscores once again how a limit on the growth of $\phi$ can 
be set by a local length scale. 

So after this analysis we can make a 
positive claim that we have provided an illustrative mathematical 
model of how in the most optimistic possible scenario (the ``Blue
Ocean"), where there is no dissipative factor to contend with, 
the growth of an industrial organization will still become terminal
in nature, simply because the environment within which the 
organization operates is conservative and has practical limitations
imposed upon it by finite but well-behaved boundary conditions. 

\section{Universal features and predictions}
\label{sec5}

Our objective in undertaking this entire study has primarily been 
to provide a mathematically viable basis for our argument that just
like any natural physical system, an industrial organization will 
also have an end to its growth. In doing so, we chose the growth 
of {\it IBM} as our case in point. We have also dwelt at length on 
what had prompted our choice, namely, the fact that {\it IBM} being
a player on the global stage, its growth figures would be free of 
local fluctuations. Besides, {\it IBM} has been in existence for 
nearly a century, and, therefore, it is ideally suited as a case
study over long time scales. 
While it has been gratifying for us to note that
we have come beguilingly close to furnishing an accurate analytical
model for industrial growth, it has also broached a very serious 
question: What can be done to defer the onset of the terminal phase
in the life of an organization? Resignation in the face of the 
inevitable cannot be an answer. One might as well argue that since
death awaits all mortals, extending the life-expectancy of individuals
through the advancement of medical science should be a colossally 
futile exercise. 

Let us, therefore, first diagnose the general symptoms of the terminal
phase in our mathematical model. The parameter, $b$, in both 
Eqs.~(\ref{stokes}) \&~(\ref{nonlin}), has been identified as being 
connected to saturation in growth. We have also argued, with the 
help of the {\it IBM} data, that Eq.~(\ref{nonlin}) gives a much 
better global model for terminal behaviour than Eq.~(\ref{stokes}). 
This will naturally suggest that the retarding factors acting against
the growth of an organization are nonlinear in character. These 
factors are all quantified through the parameter $\alpha$ in 
Eq.~(\ref{nonlin}). We have already shown in 
Figs.~\ref{f1} \&~\ref{f6} that reducing $b$ and $\alpha$, respectively, 
will be conducive to lasting growth. So, taken together, a successful 
management strategy for feasible long-term growth will entail tuning 
both $b$ and $\alpha$ periodically (alignment monitoring, so to speak)
to reduce their negative influence. 
The ``Blue Ocean" strategy is certainly a practical and effective 
step in that direction, but our mathematical modelling cautions us
against placing our absolute trust in it. 

A more comprehensive approach should necessitate making precise 
numerical estimates in relation to the various retarding parameters 
in our model. 
This can only be done empirically and on a sufficiently extensive scale. 
The rewards of this pain-staking exercise, however, could be 
far-reaching as well as revealing. As an example of this method, 
we found that the growth curve of {\it IBM} started straggling 
noticeably behind a purely exponential growth pattern around the 
fortieth year of the company, since its inception under its presently
well-recognized brand name. While this indicated the dissipative 
influence of nonlinear factors in a mathematical sense, from a 
more practical point of view in identifying these factors, it 
also pointed to a possible connection between the slowing down 
in growth and a global expansion overseas. We believe that much 
more useful information could be gleaned if similar analytical 
methods were to be applied to study the growth behaviour of other
industrial organizations. This could enable us to identify the 
common denominators active behind terminal growth.  

All the same, a precise diagnosis of the general malaise will only 
amount to addressing a part of the question of how to defer the 
onset of the terminal phase. The more important part would be to
define and present a solution. Admittedly, it will never be an 
easy task to devise an effective strategy, based on quantitative 
measures, to a sustain a steady momentum in industrial growth. It
will be even more difficult to make such a strategy successfully 
applicable in all cases. 

To elucidate the last point further, let us consider the background 
of {\it IBM}, on which our case study has been based. 
It is a company that has its primary base in the US, and is a
key player in the global computer industry. In fact {\it IBM} is
important enough to be a representative organization of the 
industrial domain to which it belongs. In consequence of this fact,
we may entertain the view that in its general import at least, 
our model for the growth of {\it IBM}
can also be extended to make a preliminary foray into studying the 
growth of other representative organizations from different industries. 
Nevertheless, in doing so we may encounter great variations in matters
related to the details. 
Just as animal species thrive and evolve in their own
ecological niches, so is it with industries. It is largely true 
that each one of them thrives in its own chosen and well-defined 
domain, catering to the specific demands of a particular market, 
and deriving their sustenance from the specialized competency of 
a trained workforce. As a result different kinds of industries may
pose its own specific array of complex problems, which could only
be understood on a case-by-case basis. While the controlling factors
in all these cases may be economic in essence, it is also very much 
possible that socio-political circumstances may leave their imprint
on the fortunes of any industry. It is certainly difficult to deny that 
the social, political and economic framework of the country of origin 
of any industrial organization will all collectively have a decisive
role to play in determining the early extent of the growth of the 
organization. For instance, it is hardly to be expected that a 
representative company of the Japanese automobiles industry
will show exactly the same features as, let 
us say, a German pharmaceuticals company. Indeed, two different 
industries from the same country could also very likely display different
traits. We are strongly of the view that all these aspects of an 
organization and its history will manifest themselves in measurable
terms through various growth-related parameters. Therefore, a study
of these parameters is of paramount importance in 
determining what the nature of
the solution should be to bring about sustained growth. 

This will also be quite useful for organizations that are still in 
the early stages of their growth. Armed with a knowledge of the general
nature of the possible difficulties that lie ahead, a company can always
apply an appropriately corrective
measure at the right juncture. As a result they 
will be capable of making a more effective implementation of future
strategies and innovative solutions for growth, all of which should 
ideally be in a state of adaptable alignment with their objectives
and core competency. 

On the other hand, companies which would already have entered a terminal
phase, might also draw some object lessons from these studies, in the 
form of proper reality checks. This will allow for a more functional
and timely redefining of fundamental objectives. The solutions which
might follow, could be varied in many unexpected ways. Rather than 
adhering to conventional modes of growth and survival, industries 
could devise ways of preserving both their existence and their 
relevance by being more integrated with the social welfare of their
markets --- in short, assist (perhaps even, in a partial measure at 
least, assume roles which hitherto were largely the preserve of the 
state) in creating an environment of overall prosperity, and a general
feeling of well-being, right alongside the generation of wealth. The
possibilities of deriving direct and indirect benefits through such 
inclusive and enlightened strategy implementation and objectives are 
myriad. Their social, political and, certainly, economic impact will 
be both lasting and salutary. We propose to take up these 
and many other related issues in subsequent studies. 

\begin{acknowledgments}
The authors are grateful to Abhik Basu, Jayanta K. Bhattacharjee,
Swati Bhattacharya, Reiner Piske and Hiren Singharay for many useful 
comments, and to 
Annup Varkey for the collection of empirical data. Gratitude is also 
to be expressed to Ajit K. Kembhavi and Subharthi Ray for their support 
and help in many respects. 
\end{acknowledgments}

\bibliography{mmr}

\end{document}